\documentclass[12pt]{article}

\usepackage{graphics}
\usepackage{epsfig}
\usepackage{amssymb,amsfonts,amsmath}

\begin{document}

\title{Opinion Particles: Classical Physics and Opinion Dynamics}
\author{Andr\'e C. R. Martins \\
NISC - EACH\\ Universidade de S\~ao Paulo, Brazil
}

\maketitle

\begin{abstract}
A model for Opinion Particles, based on Bayesian-inspired models of Opinion Dynamics such as the CODA model is presented. By extending the discrete time characteristic of those models to continuous time, a theory for the movement of opinion particles is obtained, based only on inference ideas. This will allow inertia to be obtained as a consequence of an extended CODA model. For the general case, we will see that the likelihoods are associated with variables such as velocity and acceleration of the particles. Newtonian forces are easily defined and the relationship between a force and the equivalent likelihood provided. The case of the harmonic oscillator is solved as an example, to illustrate clearly the relationship between Opinion Particles and Mechanics. Finally, possible paths to apply these results to General Relativity are debated.
\end{abstract}

%\keywords{CODA | Opinion Dynamics| Newtonian Mechanics | General Relativity | Mach principle}

\section{Introduction}

The problem with integrating Quantum Mechanics and General Relativity is so widely known that no introduction to the theme is actually required. Both theories are very successful in the cases where we know they must be applied. Yet, no agreement has been reached by the scientific community on a good model for Quantum Gravity, despite the fact that efforts do exist, as, per example, String Theory \cite{beckeretal07a} or Loop Quantum Gravity \cite{thiemann07a}. A framework that would allow both Quantum Mechanics and General Relativity to be obtained from the same basic principles could, in principle, provide a important new development in the efforts to understand how both theories could be formulated in an unifying way. If that framework were to be based on logical and informational principles, that would be a very nice bonus.

Efforts to understand Physics from an inference point of view are not new and have provided some very interesting results \cite{jaynes03,caticha04a,catichapreuss04a,catichagiffin07a,goyal12a}. They use entropic principles based on the information one receives from experiments and are compatible with Bayesian Statistics, in the cases where the information comes as data. It is interesting to notice that Bayesian Statistics can actually be seen as an extension of Classical Logic where statements are not just true or false, but can have different amounts of plausibility assigned to them, according to what is currently known \cite{cox61a,jaynes03}. That means that entropic and informational approaches use very basic first principles of reasoning. While these are interesting advances, they tend to apply inference principles only to understand how an observer would interpret the results of an experiment.

In this paper, Bayesian Logic will be shown to possibly allow for a deeper understanding of theoretical Physics than that allowed by just applying it to the inferential process. The idea of the paper is to provide initial ideas in that direction and to present far more questions than answers, while also illustrating how the general principles presented herein can be applied. This will be accomplished by expanding on existing models from Opinion Dynamics.

In Opinion Dynamics problems \cite{galametal82,galammoscovici91,sznajd00,deffuantetal00,hegselmannkrause02,martins08a,galam12a}, a framework based on Bayesian methods has been proposed \cite{martins12b} as an extension of the ideas presented in the Continuous Opinions and Discrete Actions (CODA) model\cite{martins08a,martins08b}. It is interesting to notice that this framework is able to generate many of the other Opinion Dynamics models, both discrete \cite{martins12a} and continuous \cite{martins08c}, as particular or limit cases. Those are the ideas that will be expanded here into a full model for opinion particles.

The paper is structured like this: First, a brief review of CODA model will be presented.  CODA is much simpler if we transform the probabilities $p$ to log-odds $\nu$ and, therefore, most of the analysis will be conducted using the variables $\nu$. In order to obtain a more realistic dynamics, I will proceed by extending CODA to continuous time, thus defining  opinion particles. These particles move as a consequence of an inferential process based on the information they receive, like, per example, the position (and, possibly, other variables) of the other particles. We will see that inertia can be obtained as a consequence of this continuous time CODA, with no need to introduce any kind of space structure. This means that 
Mach's principle \cite{mach88a,sciama53a,dicke63a,barbour10a,essen13a} can be applied to Newtonian Mechanics, as inertia becomes a consequence of interaction, with no need for introducing privileged inertial frames.

The relationship between likelihoods and the forces that act on the opinion particles will be obtained and we will see how the harmonic oscillator can be described in this framework. Through an inference where the full information about the location of the other particles is used, I will show that the results of the harmonic oscillator can be obtained from Beta function likelihoods.

Finally, the problem of the geometric structure of the space of $\nu$ coordinates will be briefly discussed. We will see it makes sense to identify $\nu$ with space-time coordinates and obtain the equation that gives us the likelihood equivalent to a parallel transport in a relativistic gravitational problem.

One warning is necessary before proceeding. While the name Opinion Particles in a natural name, given they are inspired in Opinion Dynamics models, it does not mean that the particles have any kind of opinion. The name just reflects the fact that, as information about other particle arrives, the first particle uses that information in some unknown process to determine its trajectory. That this use can be compatible with principles of rationality is an interesting fact, with no explanation or speculation attached to it.

\section{CODA and Inertia}\label{sec:coda}

\subsection{CODA}

CODA model \cite{martins08a,martins08b} was developed to explore how human agents could influence each other, changing their opinions when observing the choices of other agents. It assumes that, in a situation where there are two possible choices (or actions), each agent assigns a fixed probability $\alpha > 0.5$ that each one of its neighbors will have chosen the best alternative (repulsive forces are also easy to introduce by allowing $\alpha<0.5$, when we have agents known as contrarians \cite{galam04,galam05,martinskuba09a}). Let the two choices be $A$ and $B$, and $p_i (t)$ be the probability agent $i$ assigns at time $t$ to the probability that $A$ is the best choice. CODA assumes a fixed likelihood $\alpha \equiv P(OA_j|A)$, representing the chance that, if $A$ is indeed the best choice, when observing agent $j$, $i$ will observe $j$ prefers $A$, indicated by $OA_j$. If we don't assume that the problem is symmetrical in relation to both choices, that is, $\alpha \equiv P(OA_j|A) \neq \beta \equiv P(OB_j|B),$
a simple use of Bayes theorem will show how $p_i(t)$ is altered.  Per example, if agent $j$ prefers $A$, we have
\begin{equation}\label{eq:bayes}
p_i(t+1|OA_j)=\frac{p(t)}{1-p(t)}\frac{\alpha}{1-\beta}.
\end{equation}

The model becomes much simpler if we use the log-odd function 
\begin{equation}
\nu \equiv \ln(\frac{p}{1-p})
\end{equation}
(where the agent index and time dependence were omitted). As $p$ exists in the interval $0\leq p \leq 1$, we have $-\infty \leq \nu \leq +\infty$. In the $\nu$ variable, a simple additive model is obtained. That is, for the symmetric case where $\alpha=\beta$,
\begin{equation}\label{eq:coda}
\nu(t+1) = \nu(t) \pm C,
\end{equation}
with the plus sign corresponding to agent $j$ preferring $A$ and the minus sign to the opposite choice and where $C=\ln\left( \frac{\alpha}{1-\alpha} \right)$ (if $\alpha \ne \beta$, the plus and minus terms just have different sizes).

 This symmetric case can be trivially normalized to $\nu^*$ so that when $A$ is observed, the agent adds $+1$ to $\nu^*$, and when $B$ is observed, $-1$ is added, but we won't normalize $\nu$ in this paper. The choice of the agent is defined as the sign of $\nu^*$, with positive signs indicating $A$ is chosen. It is important to notice that Equation \ref{eq:coda} is actually general, even if we assume different likelihoods. The expression for $C$ can and will change, however, since $C$ is a simple constant due to the use of the simple likelihood of CODA model. 

\subsection{Continuous time}

CODA model is defined in discrete time. Each observation of new data is a discrete event and, from an inference point of view, it seems to make no sense to talk about changes between the discrete observations. Of course, this does not prevent us from extending the model to continuous time, from a mathematical point of view. One might, if desired, think of it as the whole information arriving in smaller bits.

Such an extension can be trivially obtained. If after $\Delta t =1$ we have $\Delta \nu = C$, we can assume a linear change in time so that $\Delta \nu = C \Delta t$. Actually, any power of $\Delta t$ would give the correct discrete limit, and we will see bellow that this simpler expression is not necessarily the best choice. 

As we take $\Delta t \rightarrow 0$, we can write tentatively
\begin{equation}\label{eq:velocity}
\frac{d\nu}{dt} = C.
\end{equation}
That is, $C$, while associated to the change of $\nu$, would play the role of velocity for the opinion particle. This role is basically correct if one looks for a function $C$ that describe the full movement. However, as we will see later, if we divide the dynamics into an inertial component and another arising from a force, Equation \ref{eq:velocity} will have to be replaced, as the force component will assume a particle starting from rest. In that case, $C$ will be associated with the acceleration, with a change depending on $\Delta t^2$, as we will see bellow.

In unidimensional CODA, supposing the choice was $A$, $C=\ln\left( \frac{\alpha}{1-\beta} \right)$ is obtained from the ratio between the probability that $A$ is chosen when $A$ is true ($\alpha$) and the probability that $A$ is chosen if $B$ is true $1-\beta$. For an opinion particle $i$ that observes the position $\nu_j$ of another particle, we could define an extension according to whether $\nu_j$ is negative or positive. 

However, if we want to apply this theory to physical problems, the choice of the zero for the $\nu$-axis should be free. While a translation in $\nu$ does correspond to a change of the probability $p$, this is a problem only if we hold to an objective meaning of $p$. That objective view is not available to each particle, though. From the point of view of an opinion particle, the central position ($p=0.5$ and $\nu=0$) is its location. That is, we can have a reference frame $\nu(i)$ for each particle where the positions of the other particles, $\nu_j(i)$, where $j=1,\cdots, N$ represents each one of particles in the universe, are measured. 

What we gain from these choices is that each particle is influenced just by the sum of of the influences to its right to the sum of the influences at its left. In agreement with Sciama \cite{sciama53a}, these inertial forces obtained from this CODA-like structure decay slower than the square of distance. Here, they actually don't decay at all. Of course, other functional forms can indeed provide an inertial effect, but it is interesting that a simple extension of CODA already does it.

\section{Opinion Particles in $d$ dimensions}

The discussion at the previous Section was valid for the continuous extension of the CODA model. But the class of possible dynamics based on probabilistic reasoning is much larger. Basically, as long as we have information arriving from some source and a likelihood about that information, this can be used to update the probabilities associated with any parameter. In the case of opinion particles, that simply means that the information will be processed, causing the particle to move. If we retain the probabilistic interpretation, that means a different evaluation about some parameter. If we are only interested in obtaining equations of movement, we have nothing more than information being used by a particle to alter its position. In any case, the general case for Opinion Dynamics models on discrete time based on Bayesian rules were presented before \cite{martins12b}. What we are interested here are in models in continuous time that can be used in a $d$-dimensional space. Therefore, at this point, we will not use some concepts like what how agent $j$ expresses its opinion. In this article, just the position, velocity and needed parameters of $j$ will be observed by particle $i$.

For that, we must return to the probabilistic representation and the Bayes Theorem. Notice that while CODA models is simpler on the $\nu$, it can just as well be defined on the probability space of the $p$ variable. This is a simple choice of variables, distributions on $\nu$ have equivalents on $p$ as well as any other parameterization we choose. In order to compare with Mechanics, $\nu$ just has the obvious advantage of existing in the range $-\infty < \mu < \infty$, making it easier to interpret and compare to Cartesian coordinates.

Just like regular space, opinions have several components, since they can be about several issues \cite{vicenteetal08b}. If those issues are independent, they can be represented by the existence of a dimensional space, where the opinion particles can move on. If we still assume the particles just decide to move whether towards $-\infty$ or towards $\infty$, we have something similar to the extended CODA model in $d$ dimensions. Assuming particle $j$ is situates at $\vec{x}(j)$

Therefore, we have a continuum of infinite possible values for $\vec{x}(j)$ as well as, possible, other parameters. Representing the set of parameters by $\varTheta $, we have, for the first coordinate $p_1(t+1)$ as a function of $p_1(t)$, represented simply as $p_1$ for simplicity sake,
\begin{equation}\label{eq:continuousbayes}
p_1(t+1)=\frac{p_1 f(\varTheta|A_1 )}{p_1 f(\varTheta|A_1 )+(1-p_1) f(\varTheta|B_1 )},
\end{equation}
where $A_1$ corresponds to the possibility that $p_1=1$ is the correct answer, and $B_1$ to $p_1 =0$. In a way, this might be considered equivalent to the Holographic Principle \cite{susskind95a}, in the sense that each position is simply a representation of the desired choices $p=0$ or $p=1$, that correspond to $\nu=-\infty$ and $\nu=+\infty$, respectively.

Converting to $\nu_1$, we have Equation \ref{eq:coda} again, except now we have one such equation for each coordinate, that is the vectorial equation
\begin{equation}\label{eq:codadimension}
\frac{d\vec{\nu}}{dt} =  \vec{C}
\end{equation}
where 
\[
\vec{C}=\ln \left( \frac{f(\vec{x}|A_1 )}{f(\vec{x}|B_1 )} \right) .
\]

This is not complete, yet, however, as there the assumption  that only $p=0$ or $p=1$ are correct values is not necessarily correct. Even for real opinions, it is perfectly reasonable to conclude that the best value for some parameter is not in one of its extremes. In the case we have a continuum of possible values for a best position ($p$ or $\nu$ or any other), we will need a probability distribution over the positions and it is this distribution that will be updated, with the sum in the denominator of Equation \ref{eq:continuousbayes} substituted by an integration over all possible values. This case will not be addressed here, since it is not necessary for the examples in this paper, but it is perfectly reasonable that it might be needed for some specific forces and interactions.

\section{Forces and opinions}\label{sec:forces}

Let's assume for now that $\nu$-space has the normal geometric structure of a flat, Newtonian space-time and that there are three components of $\nu$, equivalent to the three Cartesian spatial directions. A similar supposition will be made in the context of space-time coordinates.

We have obtained so far an equation for the velocities of the opinion particles (Equation \ref{eq:velocity}) and we have discussed how we can explain inertia in a straight-forward way. To compare the movements of the opinion particles with those of Newtonian particles we need to calculate the likelihoods that match known forces. Notice that $C$ is not actually a likelihood, but the logarithm of the ratio of two likelihoods (or, if the likelihoods are defined directly on $\nu$, the ratio of two likelihods).  The equivalence to a force can be achieved by derivating Equation \ref{eq:velocity}, to obtain the acceleration of the particle. Since $C$, in general, depends on the relative position of the other particles, it will change as the particle moves, the opinion particle will have a non-null acceleration. And, given we define the particle mass in a consistent way, we can derive Equation \ref{eq:velocity} in order to obtain a relation between a force $F$ known from Newtonian Mechanics and its associated likelihood. If $C$ is associated with the velocity of the particle, this is trivially given by, assuming a constant mass $m$,
\begin{equation}\label{eq:force}
F=m\frac{dC}{dt}.
\end{equation}
Integrating the expression provides us with the likelihood that is equivalent to any force up to a constant term. We will see bellow that, in several cases, $C$ might be better identified with a force instead of the velocity. In those cases, the Equation \ref{eq:force} is obviously not valid. Should $C$ be associated with $\Delta t^2$, it will play the role of an acceleration and its relation to the force is far more obvious.

\section{Velocities, inertia, and trust effects}\label{sec:trust}

One thing to notice is that, if $C$ depended only on the relative position of other particles, a given particle would have its trajectory from a fixed point completely defined, with no possibility to have it, while at that point, different velocities. This holds also for inertial forces and it would make opinion particles behave in a way that is not compatible with the movement of physical particles. While this is not a problem for Opinion Dynamics problems, it would limit the applicability of the concept. Therefore, $C$ must have other types of dependencies, so that different velocities are allowed at a point.

One possible way yo address the initial velocity problem we can introduce trust between the particles \cite{martins08c,martins13a}. Basically, this might be seen as equivalent to the problem of model choice in a Bayesian context \cite{ohagan94a}, where the likelihood is obtained from a sum of at least two different probability distributions. The term sum, of course, can mean, in the last sentence, an integration over the values of a parameter. If one of the distributions is assumed to be not related to a best position for the particle, as non-informative as possible, that part of the distribution will not contribute to the movement. In a one-dimensional model with two distributions \cite{martins08c}, if the informative distribution has a probability $q$ of being the best model, the non-informative one will have a probability $1-q$ associated to it. This means that the movement of the particle will be diminished by a factor $q<1$ when compared with the model with no non-informative part. Of course, as the particle moves, $q$ may also be updated and, as such, it does not have to be constant over the movement. This means $q$ can be seen as a measure of how much the information arriving from the other particle is considered trustworthy or useless. This can cause different velocities in any number of $d$ dimensions, since $q$ can be a function of the angles in the $d-1$ sphere. 

In terms of implementation, however, the problem is much simpler, not requiring a full model where distributions are specified and the hyper-parameter for the probability of each distribution constantly updated. What happens is that we already know the answer. The inertia component $C_I$ of any complete particle description should be
\begin{equation}\label{cinertia}
\vec{C}_I(t+dt)=\vec{v}(t),
\end{equation}
where $\vec{C}_I$ stands for the inertial likelihood and $\vec{v}(t)$ is the velocity of opinion particle at time $t$.

\section{Harmonic Oscillator}\label{sec:harmonic}

As an example of the application of the concept of using likelihoods and opinion particles to describe Newtonian particles, let's see how we can obtain an harmonic oscillator. Equation \ref{eq:force} seems to provide a direct way to compute a function $C$ given a known force $F$, in this case, $F=-kx$. Basically, we can use Newton laws to obtain how the force evolves in time and integrate Equation \ref{eq:force} in order to obtain a total function of the likelihoods given by $C \propto \sqrt{1-x^2}$. While essentially correct, this solution misses some interesting aspects of using opinion particles.

By integrating the force using the actual movement of the particle, both the effects of inertia and the force get mixed in the function $C$. However, in order to extend the ideas presented herein to General Relativity, it would be useful to separate the inertial likelihood $C_I$ from the likelihood associated to the force, $C_F$. 

In order to obtain the movement of a particle under an harmonic oscillator force, let's return to what happens in small time increments, from $t_1$ to $t_2=t_1 + \Delta t$. In this case, we can rewrite Equation \ref{cinertia} as
\[
C_I(t_2) = v(t_1)= \frac{x(t_1)-x(t_0)}{\Delta t},
\]
where $t_0 = t_1 - \Delta t$ is the instant before $t_1$. Therefore, the change in position from the inertial term is simply given by 
\begin{equation}\label{eq:inertiamovement}
x(t_2)=x(t_1)+v(t_1) \Delta t = x(t_1) + [x(t_1)-x(t_0)].
\end{equation}
If $C_I$ were the only thing affecting the particle, its influence would be constant and, therefore, the opinion particle would trivially follow a straight line.

It remains to introduce the effect of the force, by using Equation \ref{eq:force}. We have that
\[
-kx(t_1) \Delta t = m \Delta C = m \left( C(t_2) - C(t_1) \right)
\]
Since $C$ is basically the velocity of the particle, we would have, in principle,
\begin{equation}\label{eq:cdiscrete}
-kx(t_1) \Delta t  = m \left( \frac{x(t_2)-x(t_1)}{\Delta t} - \frac{x(t_1)-x(t_0)}{\Delta t} \right).
\end{equation}
If we were to solve this, however, we would be back to using the full movement of the particle and not just the effect of the force. If we mean to have just the effect of $C_F$ from Equation \ref{eq:cdiscrete}, we must assume that the particle was initially at rest, that is $ \frac{x(t_1)-x(t_0)}{\Delta t} =0$. That way, we have
\begin{equation}\label{eq:fmovement}
x(t_2) = x(t_1) -\frac{k}{m}x(t_1) (\Delta t)^2. 
\end{equation}

Combining the movements from inertia and the force, we have
\begin{equation}\label{eq:totalmovement}
x(t_2) = x(t_1) -\frac{k}{m}x(t_1) (\Delta t)^2+ [x(t_1)-x(t_0)]. 
\end{equation}

\begin{figure}\label{fig:harmonic}
\centering
\epsfig{file=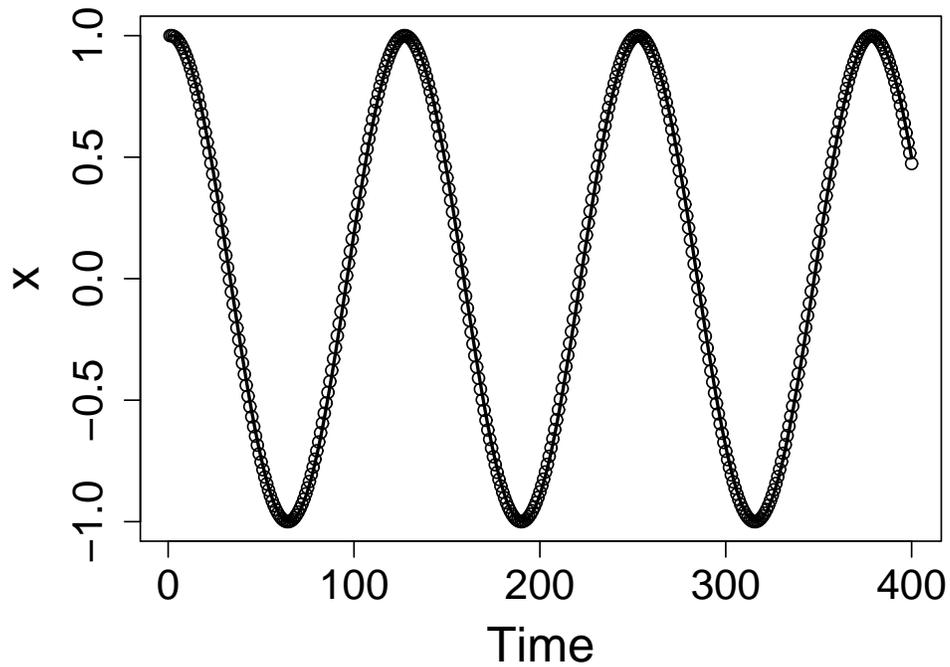,width=1.0\linewidth,clip=}
\caption{Time evolution of the discrete harmonic oscillator, given by Equation \ref{eq:totalmovement} compared to the exact solution of the same problem. In the figure, the discrete points are represented as circles and the exact solution is given by the line. In the case shown, $\Delta t=0.05$ and $\omega =1$, while time in the figure is given as multiples of $\Delta t$.}
\end{figure}

Figure 1 shows the time evolution obtained from Equation \ref{eq:totalmovement} as circles, compared to the exact solution of the unidimensional harmonic oscillator, represented as the line. We can see clearly that, despite its unconventional form, Equation \ref{eq:totalmovement} does provide the correct dynamics for the problem.

\section{Discrete Choices with Continuous Verbalization}~\label{sec:concoda}

Finally, in order to better explore the formalism and also in order to make some of its properties clearer, a variant of the CODA model will be discussed where the communication between the agents is not a discrete spin value, meaning the observed particle is at what side of the observing one.  Instead, the full probability $p_j$ particle $j$ assigns to the possibility (measured in the reference frame of $i$) that the right choice is $x=+1$ will be used in the inference made by the particle $i$. Notice that the fact that the communication is continuous does not imply that $p$ should be. We still have a problem with only two possible real choices $p=-1$ or $p=+1$. However, the continuous probabilistic value is the communicated information. This distinction is a very important albeit neglected one. In continuous opinion models, it is usually assumed that both the communication and the decision are continuous, but that doesn't have to be the case.

As the communication phase in the framework was changed, we need now a new likelihood, that neighbor agent $j$ will issue the value $p_j$ in the case where $p=-1$ and in the case where $p=+1$, that is, functions $f(p_j|A)$ and $f(p_j|B)$. Since all values for $p_i$ are limited to $0\leq p_i \leq 1$, the simplest choice is to take Beta distributions $Be(p_j|\alpha,\beta)$ as priors.

\[
Be(p_j|\alpha,\beta)=\frac{1}{N(\alpha,\beta)}p_{j}^{\alpha-1}(1-p_{j})^{\beta-1}
\]
where $N(\alpha,\beta)$ is obtained from Gamma functions by
\[
N(\alpha,\beta)=\frac{\Gamma(\alpha)\Gamma(\beta)}{\Gamma(\alpha+\beta))}.
\]

As the Beta function is symmetric in $\alpha$ e $\beta$, if we want to keep the symmetry between $p=0$ and $p=1$, we must have for the likelihoods that, if $f(p_j|x=1)=B(\alpha,\beta)$, then $f(p_j|x=-1)=B(\beta,\alpha)$.  By applying the Bayes Theorem to this problem, agent $i$, when observing $p_j$,  will update $p_i$ to $p_i(t+1)$, 
\begin{equation}\label{eq:continuouscoda}
	p_i(t+1)=\frac{p_i p_j^{\alpha-1}(1-p_j)^{\beta-1}}{p_i p_j^{\alpha-1}(1-p_j)^{\beta-1} + (1-p_i)p_j^{\beta-1}(1-p_j)^{\alpha-1}}.
\end{equation}
If we adopt the same transformation of variables as in CODA model and calculate $\nu_i$ we will see that the denominators cancel out and we have that
\begin{equation}\label{eq:logoddcontinuouscoda}
\ln \left( \frac{p_i(t+1)}{1-p_i(t+1)} \right)  =\ln \left( \frac{p_i(t)}{1-p_i(t)} \right) +\ln \left[ \left(\frac{p_j(t)}{1-p_j(t)} \right)^{\alpha-\beta} \right]
\end{equation}

Equation \ref{eq:logoddcontinuouscoda} can be rewritten in a more elegant fashion as
\begin{equation}\label{eq:logoddcontinuouscodanu}
\nu_i(t+1)=\nu_i(t)+(\alpha-\beta) \nu_j
\end{equation}
This is similar to the CODA dynamics, except that now, at each step, instead of adding a term that is constant in size and only varies in sign, we add a term proportional to the log-odds of the opinion of the neighbor.

However, we must also remember that, unlike the CODA case, that led us to inertia, the objective here is to obtain an equivalent to a force component. In Equation \ref{eq:fmovement}, we see that a force component enters as a term with $\Delta t^2$, while, if we repeat the argument that associated $C$ with velocity, we will just have a $\Delta t$. More importantly, very simple simulations of the limit $\Delta t \rightarrow 0$ show that a term  $\Delta t^2$ is required in order for the dynamics to have a proper limit. Per example, if, in Equation \ref{eq:totalmovement}, the force term had just a linear dependence on $\Delta t$, the system would still oscillate in an apparently correct way for a fixed $\Delta t$, but with a frequency that would grow as $\Delta t \rightarrow 0$.

Of course, that might not seem reason enough. We certainly need to understand why we can use a quadratic instead of a linear term, since $C$ seemed more easily associated with velocities. But we must remember that we have divided the dynamics in two parts, a inertial part and one associated with forces. The task of keeping the velocities the same as before is performed by $C_I$. And, as we have seen for the harmonic oscillator, the way $C_F$ works is by providing the same change in position the particle would have if it were at rest. That is, expanding the movement in a power series, the linear term, giving the velocity due to $C_F$ must indeed be null. The first contribution, therefore, is the acceleration, associated with $\Delta t^2$.

That means we will have, from Equation \ref{eq:logoddcontinuouscodanu} plus the inertial term, when we start heading towards continuous time, the following equation for the movement under the influence of Beta likelihoods:
\begin{equation}
\nu_i(t+1) = \nu_i(t) +(\alpha-\beta) \nu_j \Delta t^2 + [\nu_i(t)-\nu_i(t-1)],
\end{equation}
which is the same as the Equation \ref{eq:totalmovement} for the harmonic oscillator with $\nu$ in the place of $x$, for a proper choice of $\alpha$ and $\beta$. The main difference is that instead of $\nu_i$ the term in  $\Delta t^2$ has $\nu_j$. If one assumes, as mentioned before, that a particle, at each instant, adjusts its coordinate system so that the frame will return to the center, that is, we have an evolving referential frame $F$ where $\nu_{i_F} (t)=0$, then $\nu_j$ is exactly the distance between $i$ and $j$. That is, in order to recover the results from Newtonian Mechanics, it seems we must have translation invariance.

It is worth remembering that the models here assume the particles are making an inference whether to move to $-\infty$ or to $+\infty$. This is not a necessary characteristic of forces acting on opinion particles, they could very well be trying to establish a value $\nu$ to move to. It is interesting, however, that, for both inertial movement and the harmonic oscillator that was not necessary.

\section{Coordinate Changes and Metrics}\label{sec:coordinates}

We have seen the first details about how opinion particles relate to Newtonian Mechanics. However, if we want the dynamics of the opinion particles to be more general, it should also provide ways to describe central theories of modern Physics, as General Relativity and Quantum Mechanics. Here, however, I will just investigate if it makes sense for a likelihood to be used in order in a way that corresponds to the geodesics in a space where points are identified by the $\nu$ variable. The objective is simply to provide some evidence that this framework could indeed be useful to describe gravity. The already observed need for a local reference frame as well as the separation of the movement in an inertial part and the part associated to other forces seem to suggest that General Relativity should be a natural way to deal with  $\nu$ coordinates or any other coordinates we might decide to use to parameterize the space-time. 

As we have discussed, $\nu$ has no predefined geometric structure and, by identifying a likelihood with movement on geodesics, we might be able to get a new way to describe Gravity that could be compatible with General Relativity. It should be noter that connections between distributions and geometrical properties and curved spaces are not new in inferential problems \cite{amarinagaoka00a}. 

Since $\nu$ coordinates have no special geometric interpretation to them, vectors in one point (per example, the velocity of the particle) have no natural way to be defined as parallels at other point. 
The aim, if we want to find how opinion particles can represent the movement of a particle in a gravitational relativistic field, is to be able to define the likelihoods in terms of the metric of the space-time. 

\subsection{Parallel Transport}

So far, we have worked only with the spatial coordinates. We need to define the model in the space-time and, for such, extending the likelihood to act on a four-vector becomes necessary. In the Newtonian models, we had time simply pass in equal infinitesimal steps. That means the quantity $C^\alpha$ will need a time component. From Equation \ref{eq:codadimension}, we can see that, in the discrete case, the purpose of $C^\alpha$ is to update the components of the position vector $\nu^\alpha$. That is, the time component for $C^\alpha$ should be the time passed between the two observations in the frame of reference of a particle moving on a geodesic between the points. This is valid for any likelihoods in a space-time. In order to differentiate the one that should be equivalent to gravity, I will denote it by $G^\alpha$ from now on.

Since $G^\alpha$ is associated with how the coordinates of a particle change in a gravitational field, it must correspond to the geodesics. That is, it must have the same effect that parallel transport has on vectors, which means that, despite its notation, it is not really a 4-vector field. This is easy to understand, since $G^\alpha$, as it was the case in Newtonian Mechanics, needs to be not just a function of the position $\nu^\alpha$ but also of the velocity the particle arrives at $\nu^\alpha$ and, therefore, might have different values at the same point for different particles.

Following the same reasoning behind Equation \ref{eq:codadimension}, we must move from the discretized probabilistic model to a continuum one. Here, however, we must replace time for an invariant quantity, therefore, we need to differentiate the position vector in relation to the distance element $ds$. In this case, Equation \ref{eq:codadimension} becomes
\begin{equation}\label{eq:codagr}
\frac{d\nu^{\alpha}}{ds} = G^{\alpha}.
\end{equation}

A relation between the log-odds space and the coordinate space-time of General Relativity is needed to proceed. Given the linear effect $G^\alpha$ already has on $\nu$ coordinates, I will assume that both coordinate systems are the same, that is that $x^\alpha=\nu^{\alpha}$. This makes calculating the field $G^\alpha$ possible for a given known gravitational problem, simply by observing that the term $\frac{d\nu^{\alpha}}{ds}$ in Equation \ref{eq:codagr} can be obtained easily from the geodesic equation of the movement of a particle (from any General Relativity textbook, as, per example,\cite{dirac96a})
\begin{equation}\label{eq:geodesic}
\frac{du^\alpha}{ds}=-\Gamma^{\alpha}_{\beta\gamma}u^\beta u^\gamma,
\end{equation}
where $u^\alpha=\frac{d\nu^\alpha}{ds}= G^{\alpha}$.

Finally, it still needs to be investigated if the continuum limit would work properly this way or would require a dependence in $ds^2$. This, however, would mean a very simple change in Equation \ref{eq:geodesic}, since it is already the second derivative of $G$ that enters the equation.

\section{Discussion}

The work presented here is expected to provide a first step towards building a complete theory for the movement of opinion particles. As an example of its applicability in physical problems, we have shown how we can describe the movement of a harmonic oscillator using this framework. In order for more applications to Physics to become real, of course, other solutions that match likelihoods and forces or metrics must be found. The exact functions that do that translation will be studied in the future. I hope the functions prove to be simple enough to allow analytical calculations, but there is no reason for that to be the case.

In any case, the possibility of defining any forces (and, therefore, potentials) and Gravitation in the same theoretical framework opens a new possible path of exploration towards building an unified description of the fundamental physical laws. This possibility alone makes the theory for the movement of opinion particles quite worth exploring. One extra advantage it already provides is in the area of Opinion Dynamics, with a way to transform discrete opinion models that need simulation techniques into differential equations that might have an analytic solution.

Of course, many problems are still open. Really describing gravity and quantizing the theory are the main ones to be tackled in the near future.

\section*{Acknowledgments}
The author would like to thank the Funda\c{c}\~ao de Amparo \`a Pesquisa do Estado de S\~ao Paulo (FAPESP) for partial support to this research under grant 2009/08186-0.

\end{document}